\documentclass[%
 reprint,
superscriptaddress,
 amsmath,amssymb,
 aps,
prl,
]{revtex4-2}

\usepackage{svg}
\usepackage{graphicx}
\usepackage{dcolumn}
\usepackage{bm}
\usepackage{xcolor}
\usepackage{float}	
\usepackage{soul}
\usepackage{url}
\usepackage{lipsum}
\usepackage{siunitx}

\DeclareSIUnit{\RPM}{\text{RPM}}
\DeclareSIUnit{\bar}{\text{bar}}
\DeclareSIUnit{\mbar}{\milli\bar}
\DeclareSIUnit{\angstrom}{\text{Å}}
\DeclareSIUnit{\atomicpercent}{\text{at}\%}
\DeclareSIUnit{\ev}{\text{eV}}
\DeclareSIUnit{\mev}{\text{meV}}
\DeclareSIUnit{\kelvin}{\text{K}}
\DeclareSIUnit{\nm}{\text{nm}}
\DeclareSIUnit{\mortsgna}{\text{Å$^{-1}$}}

\begin{document}


\title{Electronic Structure and Resonant Circular Dichroism of La$_{0.7}$Sr$_{0.3}$MnO$_3$ from Soft X-ray Angle-Resolved Photoemission}

\author {\O yvind~Finnseth}
\affiliation {Department of Materials Science and Engineering, Norwegian University of Science and Technology, 7491 Trondheim, Norway}

\author{Damian Brzozowski}
\affiliation {Department of Materials Science and Engineering, Norwegian University of Science and Technology, 7491 Trondheim, Norway}

\author {Anders~Christian~Mathisen}
\affiliation {Center for Quantum Spintronics, Department of Physics, Norwegian University of Science and Technology, 7491 Trondheim, Norway}

\author {Stefanie~Suzanne~Brinkman}
\affiliation {Center for Quantum Spintronics, Department of Physics, Norwegian University of Science and Technology, 7491 Trondheim, Norway}

\author {Xin~Liang~Tan}
\affiliation {Center for Quantum Spintronics, Department of Physics, Norwegian University of Science and Technology, 7491 Trondheim, Norway}

\author {Fabian Göhler}
\affiliation {Center for Quantum Spintronics, Department of Physics, Norwegian University of Science and Technology, 7491 Trondheim, Norway}

\author{Benjamin A. D. Williamson}
\affiliation {Department of Materials Science and Engineering, Norwegian University of Science and Technology, 7491 Trondheim, Norway}

\author{Kristoffer Eggestad}
\affiliation {Department of Materials Science and Engineering, Norwegian University of Science and Technology, 7491 Trondheim, Norway}

\author{Meng-Jie Huang}
\affiliation{Ruprecht Haensel Laboratory, DESY, 22607 Hamburg, Germany}

\author{Jens Buck}
\affiliation{Institut f\"{u}r Experimentelle und Angewandte Physik, Christian-Albrechts-Universit\"{a}t zu Kiel, 24098 Kiel, Germany}
\affiliation{Ruprecht Haensel Laboratory, DESY, 22607 Hamburg, Germany}

\author{Moritz Hoesch}
\affiliation{Deutsches Elektronen-Synchrotron DESY, Notkestra{\ss}e 85, 22607 Hamburg, Germany}

\author{Kai Rossnagel}
\affiliation{Institute of Experimental and Applied Physics, Kiel University, 24098 Kiel, Germany}
\affiliation{Ruprecht Haensel Laboratory, DESY, 22607 Hamburg, Germany}

\author{Sverre M. Selbach}
\affiliation {Department of Materials Science and Engineering, Norwegian University of Science and Technology, 7491 Trondheim, Norway}

\author {Hendrik~Bentmann}
\affiliation {Center for Quantum Spintronics, Department of Physics, Norwegian University of Science and Technology, 7491 Trondheim, Norway}

\author{Ingrid Hallsteinsen}
\affiliation {Department of Materials Science and Engineering, Norwegian University of Science and Technology, 7491 Trondheim, Norway}


\begin{abstract} 

Coupling between spin, orbital, charge, and lattice degrees of freedom in transition‑metal oxides produces a variety of electronic and magnetic phenomena of importance for future technologies.
Here, we explore the electronic band structure of a (111)-oriented La$_{0.7}$Sr$_{0.3}$MnO$_3$ thin film through soft X-ray angle-resolved photoemission spectroscopy (ARPES).
The measurements agree with the electronic band structure calculated with density functional theory using Hubbard $U$ correction. 
Furthermore, we probe the circular dichroism in ARPES, and observe a pronounced momentum-resolved magnetic circular dichroism in resonant photoemission from the Mn $L$-edge. The approach combines the
momentum- and spin-selectivity of ARPES and X-ray magnetic circular dichroism, respectively, which could provide a useful approach for the study of unconventional magnetism.

\end{abstract}

\maketitle 


The coupling between charge, spin, orbital, and lattice degrees of freedom in transition-metal oxides (TMO) \cite{Tokura2000OrbitalOxides, Chen2021Spin-chargeOxides, Takayama2021SpinOrbit-EntangledCompounds} gives rise to physical properties of scientific and technological interest, such as colossal magnetoresistance \cite{Kimura1996InterplaneCrystal} and multiferroicity \cite{Ahn2021DesigningMaterials}.
%
Through these degrees of freedom, tunable functionality may be achieved.
Prominent examples include improving the transport properties of TMO-based transistors by interfacing with diamond \cite{Yin2018EnhancedDiamond}, or amplifying magnetic frustration in La$_2$NiO$_4$ compressively strained by the substrate \cite{Biao2024Strain-tunedLa2NiO4}.
%
Specifically, recent work has focused on perovskite thin films grown in unconventional lattice orientations, exhibiting a wide range of properties not present in their bulk counterparts.
These include enhanced interfacial conductivity \cite{Herranz2012HighInterfaces}, orbital patterning \cite{Middey2016MottNickelate}, and emergent magnetism \cite{Gibert2012ExchangeSuperlattices}.


In this context a well-studied TMO is  La$_{0.7}$Sr$_{0.3}$MnO$_3$ (LSMO) with its room-temperature ferromagnetism, half-metallicity \cite{Park1998DirectFerromagnet}, and colossal magnetoresistivity \cite{Trajanovic1996GrowthSilicon}, which have been studied for a variety of applications, such as magnetic tunnel junctions \cite{Samet2003EELSJunctions, Pawlak2022Room-TemperatureJunction}, freestanding films \cite{Brand2025EpitaxialOxides, Du2024EfficientFilms}, and magnetoelectric memory \cite{Li2023EffectDevices}.
In the case of LSMO thin films, the majority of the work has been focused on films in the pseudocubic (001) orientation, see, for example Refs. \cite{Boschker2011OptimizedCharacteristics, Cesaria2011LSMOSpintronics}.
However, films in the (111) orientation exhibit physical properties that are distinct from those in the (001) orientation.
For instance, the magnetic anisotropy of (111)-oriented films is sixfold in-plane locally, with a prevailing isotropic macroscopic response \cite{Hallsteinsen2015CrystallineFilms}, as compared to the biaxial in-plane magnetization of the (001)-oriented films \cite{Berndt2000MagneticFilms}.
Additionally, the magnetic domain sizes in (111)-oriented films is significantly larger than those found in the (001) orientation \cite{Hallsteinsen2015CrystallineFilms}.
Furthermore, interfacing LSMO with LaFeO$_3$ has been shown to induce ferromagnetic behavior in LaFeO$_3$, with the induced magnetic moment an order of magnitude larger in the (111) orientation than in the (001) orientation \cite{Hallsteinsen2016ConcurrentLa0.7Sr0.3MnO3/LaFeO3, Bruno2015InsightMapping}.
Finally, the insulating and magnetically inactive layer known to form at interfaces or surfaces in films of LSMO, often called the ``dead layer", is absent in (111)-oriented superlattices of LSMO and SrTiO$_3$ \cite{Guo2018RemovalDesign}. 
Despite this importance, experimental investigations of the electronic structure of (111)-oriented LSMO are lacking, with previous work only considering the (001) orientation \cite{Lev2015FermiMagnetoresistance, Horiba2016IsotropicLa0.6Sr0.4MnO3, Chikamatsu2006BandSpectroscopy, Krempasky2008EffectsO3}.

In this work, we report soft X-ray angle-resolved photoemission (ARPES) measurements of (111)-oriented LSMO thin films, exploiting the increased bulk sensitivity at soft X-ray energies to probe the three-dimensional bulk electronic structure of LSMO. We support the interpretation of our experimental results by first-principles DFT+$U$ calculations.
Furthermore, we study circular dichroism in ARPES for resonant excitation at the Mn $L$-edge and for non-resonant excitation. A pronounced magnetic CD (MCD) is observed in resonance, reflecting the ferromagnetic order of LSMO. At off-resonant energies, the MCD is found to be negligible. 

\section{Experimental Details}

A thin film of \qty{6.2}{\nm} of La$_{0.7}$Sr$_{0.3}$MnO$_3$ was grown epitaxially on a (111)-oriented SrTiO$_3$ substrate by pulsed laser deposition.
Samples were transferred under atmospheric conditions from the growth setup to the ultra-high vacuum  photoemission setup. In order to obtain a chemically clean and ordered surface prior to the photoemission experiment, the sample was annealed at a temperature of $\sim$ \qty{640}{\kelvin} in O$_2$ gas with a partial pressure of \qty{1e-5}{\mbar}, following previous work on the surface preparation of LSMO \cite{Monsen2012SurfaceStudy}. 
%
Successful surface preparation was confirmed by X-ray photoelectron spectroscopy and low-energy electron diffraction, indicating a clean, well-ordered surface after the preparation procedure. For details on sample growth and surface preparation, see Supplemental Material \cite{supplemental}.

Soft X-ray ARPES measurements were conducted using the ASPHERE III end station at the variable polarization XUV beamline P04 of the PETRA III storage ring at DESY (Hamburg, Germany) \cite{Viefhaus2013ThePerformance}.
ARPES data were collected at sample temperatures of approximately \qty{30}{\kelvin} and pressures below \qty{5e-10}{\mbar}. The energy resolution of the ARPES measurements was set to \qty{75}{\mev}. We used the angular deflection mode of the SCIENTA DA30-L analyzer, allowing for acquisition of angular distributions in a fixed sample geometry. Light incidence was within the analyzer slit plane at an angle of 20$^{\circ}$ to the sample surface [see Fig. \ref{fig:xmcd_valence}(a)]. 
For all measurements, both left and right circularly polarized light was used. Unless otherwise stated, the results presented are obtained by taking the sum of the two different polarizations such that $I = I_L+I_R$.
To mitigate effects of beam-induced damage to the thin film, the sample was moved so that a new spot was illuminated at 5 minute intervals.

\section{Computational details}

Density functional theory (DFT) calculations were performed using VASP \cite{Kresse1993AbMetals, Kresse1996EfficientSet, Kresse1996EfficiencySet} to obtain the electronic band structure of LSMO. 
The effect of mixed occupancy of La and Sr was accounted for with supercells generated using the special quasirandom structure (SQS) method \cite{Zunger1990SpecialStructures} as implemented in the \texttt{icet} Python package \cite{Angqvist2019ICETExpansions}. As a starting point for the supercell generation, a cubic LaMnO$_3$ primitive unit cell was expanded into a $4\times2\times2$ supercell with 80 atoms, out of which 11 La and 5 Sr, giving a stoichiometry of La$_{0.69}$Sr$_{0.31}$MnO$_3$.

The supercell geometry, including shape and volume, was optimized using a plane wave cutoff energy of 750 eV and a $\Gamma$-centered $k$-point grid of $2\times4\times4$ such that the forces acting on the ions were below 0.01 eV Å$^{-1}$.
The PBEsol functional \cite{Perdew2008RestoringSurfaces} was used along with Dudarev's \cite{Dudarev1998Electron-energy-lossStudy} implementation of Hubbard $U$ correction with $U$ = 3 eV on Mn 3d \cite{Hmok2020EffectCalculations, Moreau2018OctahedralHeterostructures}. PBEsol+$U$ has been shown to yield good agreement with the structural properties of LaMnO$_3$ \cite{Mellan2015ImportanceLaMnO3, Lee2025RevisitingStudy, Gavin2017ModellingLaMnO3} and SrMnO$_3$ \cite{Zhu2020MagneticSrMnO3, Edstrom2018First-principles-basedSrMnO3, Marthinsen2016CouplingPerovskites}.
The La(5s$^2$5p$^6$5d$^1$6s$^2$), Sr(4s$^2$4p$^6$5s$^2$), Mn(3s$^2$3p$^6$3d$^6$4s$^1$) and O(2s$^2$2p$^4$) states were treated as valence electrons, while the interaction with the core states was accounted for using the projector-augmented wave method (PAW) \cite{Kresse1999FromMethod}.


To address band folding induced by supercell calculations, the \texttt{easyunfold} Python package \cite{Zhu2024Easyunfold:Structures} was used to obtain the unfolded band structure.
The bands were unfolded onto a cubic primitive cell, and are thus expressed in a simple cubic coordinate system.
In addition to obtaining unfolded band structures, this approach yields results with proper spectral weight, facilitating more direct comparison to ARPES measurements \cite{Ku2010UnfoldingStructures}, as previously shown in e.g. Refs. \cite{Rubel2023BandSpectra, Iwata2017MiningMethodology}.

Fermi surfaces were obtained by calculating band structures using k-point grids with a spacing of approximately \qty{0.02}{\mortsgna} in the reciprocal planes of interest.
The resulting band structures were unfolded onto the primitive cubic cell, and subsequently interpolated to construct continuous surface representations. 
The Fermi level in the DFT calculations was shifted 
to match the experimental results.

\section{Results and Discussion}

\subsection{Electronic structure of LSMO(111)}

\begin{figure}
    \centering
    \includegraphics[width = \linewidth]{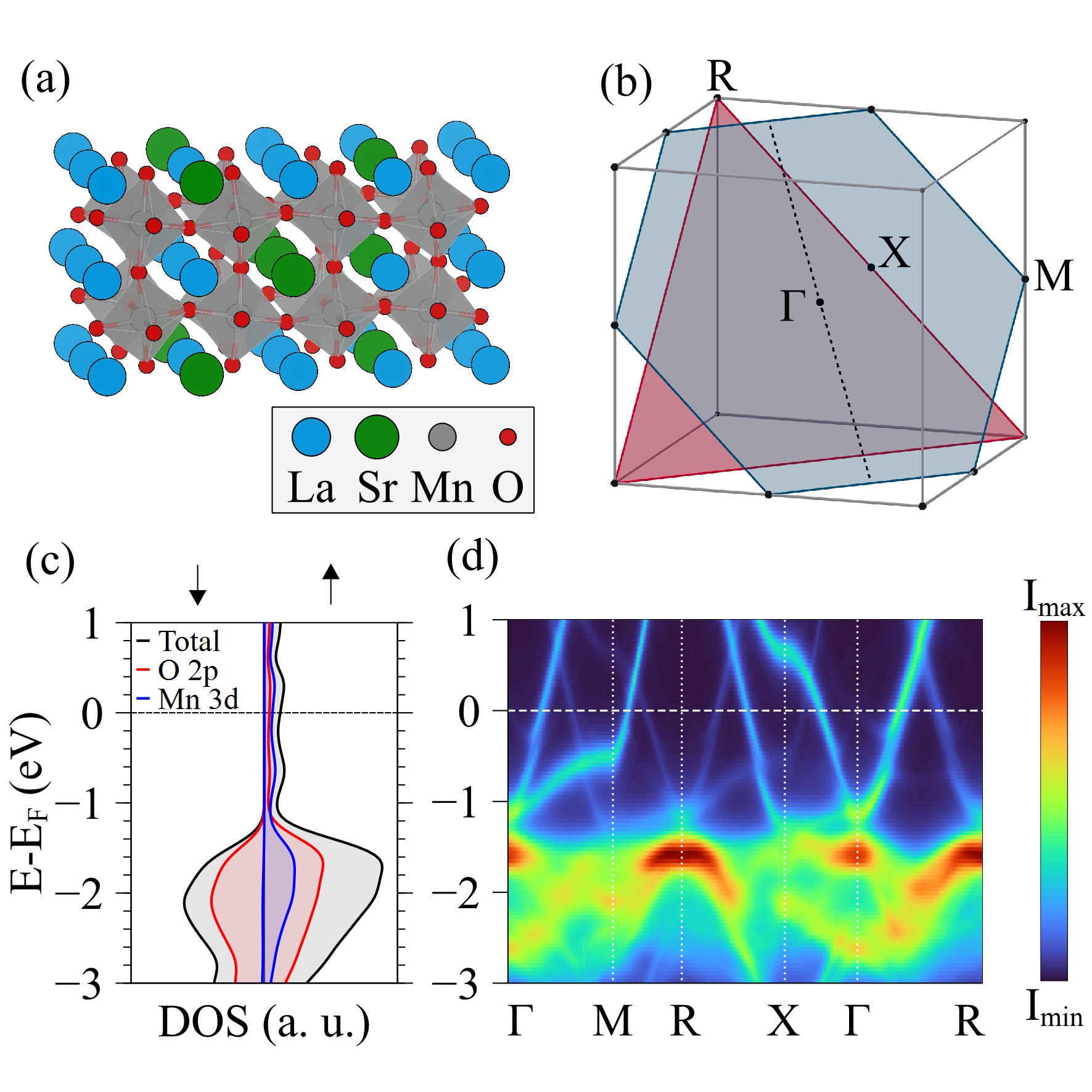}
    \caption{
    (a) Supercell used in the DFT+$U$ calculations. 
    (b) Diagram of the cubic Brillouin zone with high symmetry points indicated.
    (c) Density of states calculation for spin-up (left) and spin-down (right) electrons. All contributions besides O 2p and Mn 3d are negligible in the plotted range.
    (d) Unfolded spin-up band structure diagram calculated by DFT+$U$.
    }
    \label{fig:scfig}
\end{figure}


First we present the DFT results on the electronic structure of LSMO.
The SQS-generated supercell is shown in Fig. \ref{fig:scfig}(a).
%
%
The calculated density of states (DOS) for this supercell, shown in Fig. \ref{fig:scfig}(c), exhibits full spin polarization at the Fermi level, as expected for a half-metallic system. 
The unfolded spin-up electronic band structure is shown in Fig. \ref{fig:scfig}(d) with high symmetry points as denoted in Fig. \ref{fig:scfig}(b). 
Most prominent are the electron-like bands around the $\Gamma$ point and the hole-like bands around the R point. There are, however, fainter hole-like states around the $\Gamma$ point, and electron-like states around the R point as well. 
These are, in agreement with previous studies \cite{Lev2015FermiMagnetoresistance, Bruno2017ElectronicARPES}, attributed to the tilting of the Mn octahedra causing a doubling of the unit cell along the [111] direction. 
This, in turn, causes backfolding between the R-X plane and the $\Gamma$-M plane [see Fig. \ref{fig:scfig}(b)]. 
%
This interpretation is strengthened by the backfolded bands being absent for calculations conducted on the same supercell with the octahedral tilting removed (see Supplemental Material \cite{supplemental}).
%


Next we provide a comparison between the experimental and calculated electronic structure, focusing first on the dependence on out-of-plane momentum along the surface normal.
In order to probe the out-of-plane momentum, a photon energy scan was conducted between energies of \qty{300}{\ev} and \qty{530}{\ev}.
The orientation of the $k_x$-axis is denoted by the dashed line in Fig. \ref{fig:scfig}(b), corresponding to the [$11\bar{2}$] direction, while the out-of-plane direction $k_z$ is along $[111]$.
For conversion from photon energy to $k_z$, the free-electron final state approximation was used with an inner potential of $V_0 = $ \qty{10}{\ev} \cite{Damascelli2004ProbingARPES}.
With this inner potential, a $\Gamma$-point is located at $k_z = $ \qty{11.2}{\mortsgna}. 

\begin{figure}
    \centering
    \includegraphics[width=\linewidth]{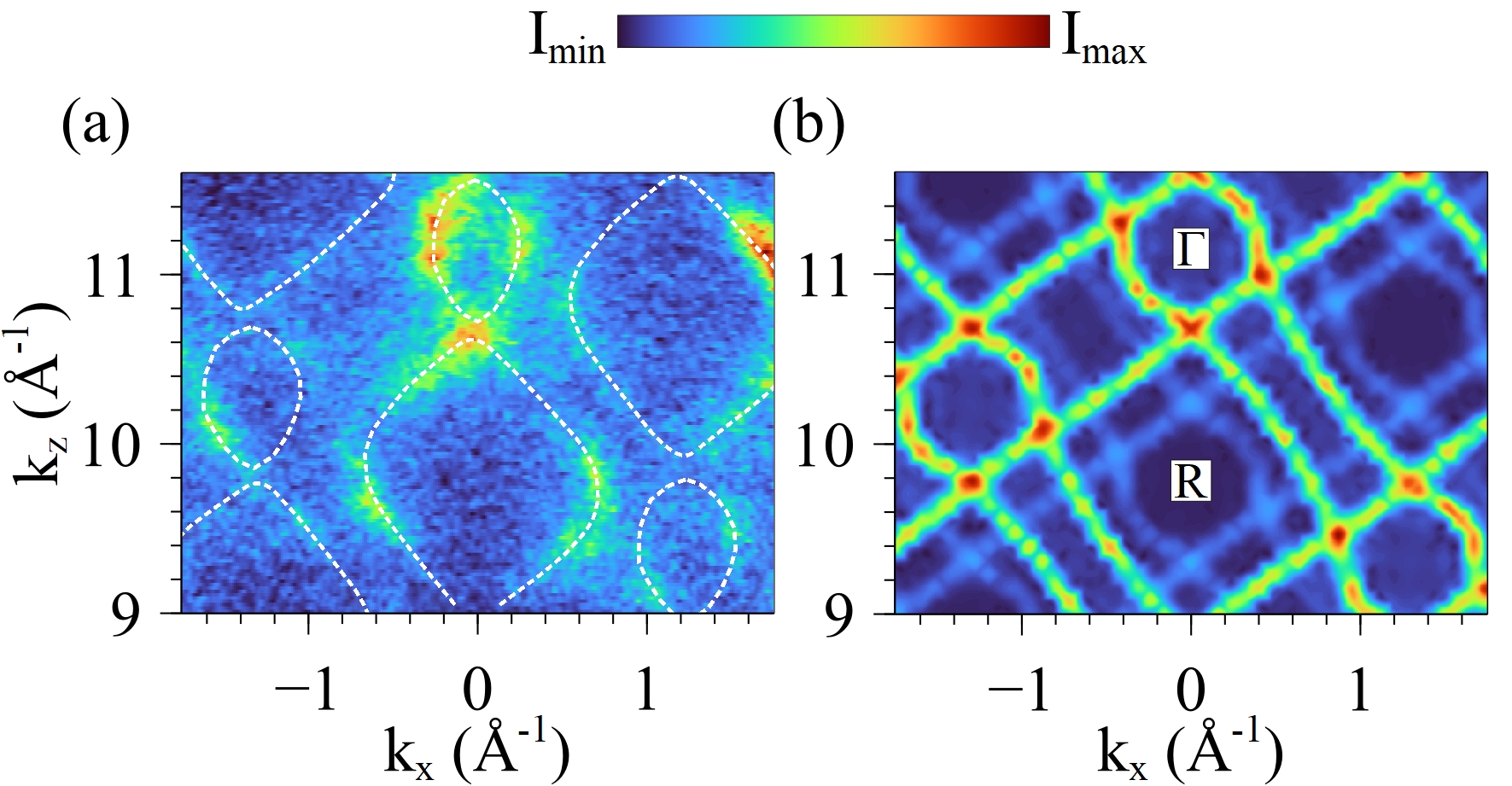}
    \caption{
    (a) ARPES momentum map at the Fermi level along the [111] (out-of-plane) direction $k_z$ and the in-plane direction $k_x$ along [$11\bar{2}$] as indicated in Fig. \ref{fig:scfig}(b). 
    The data was obtained by varying the photon energy from $h\nu =$~\qty{300}{\ev} to \qty{530}{\ev}. 
    Two main features are observed: a large pocket around the R point, and a smaller pocket around the $\Gamma$ point. Traces of these features are added as guides to the eye.
    (b) DFT+$U$ calculated Fermi surface along the same cut through the Brillouin zone as in (a).
    }
    \label{fig:kz}
\end{figure}

The resulting cut through the Fermi surface, shown in Fig. \ref{fig:kz}(a), includes a larger hole-like pocket around the R point, and a smaller electron-like pocket around the $\Gamma$ point, as expected from our band calculation in Fig. \ref{fig:scfig}(d).
The measured Fermi surface is in agreement with a calculation of the spin-up Fermi surface in the same cut through the Brillouin zone, shown in Fig. \ref{fig:kz}(b). 
As in Fig. \ref{fig:scfig}(d), the unit cell doubling causes backfolding of bands between R and $\Gamma$. The backfolded bands are however not discernible in the experimental Fermi surface. 
Considering the low spectral weight the DFT calculations predict for these states, it is likely that they are too faint to resolve in the ARPES measurements. On the other hand it could indicate that the octahedral tilting in the LSMO thin film deviates from expected bulk pattern. 
This interpretation may be supported by reports of a region with varying A-cation distance persisting several nanometers from the interface in LSMO films grown on (111)-oriented SrTiO$_3$ \cite{Nord2017Atomap:Fitting}, which could in turn influence the tilting pattern.
Precisely measuring the octahedral tilt in thin films is, in general, challenging, but has recently been demonstrated for (001)-oriented LSMO films on various substrates \cite{Dey2025AnInterfaces}.

\begin{figure*}[hbpt!]
    \centering
    \includegraphics[width = \linewidth]{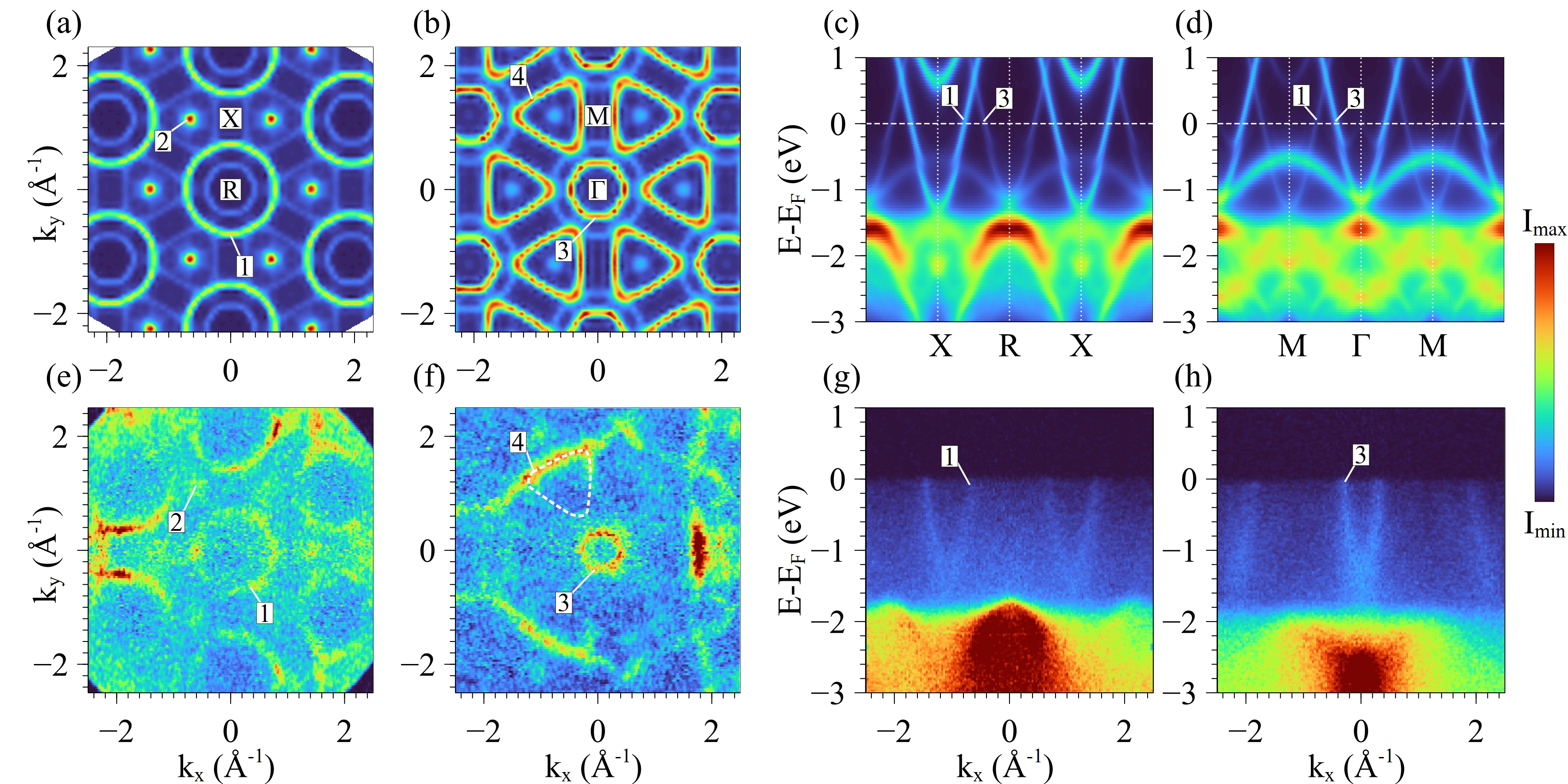}
    \caption{
    Fermi surface cuts calculated by DFT+$U$for the (a) R-X and (b) $\Gamma$-M high symmetry planes [cf. Fig. \ref{fig:scfig}(b)].
    The labels (1-4) denote the most prominent features in each of the Fermi surface cuts.
    Calculated valence band dispersions along the (c) R-X and (d) $\Gamma$-M directions.
    Bands crossing the Fermi level are marked according to the labels in (a) and (b).
    ARPES momentum maps at the Fermi level taken in the (e) R-X high symmetry plane, obtained with a photon energy of $h\nu = $ \qty{369}{\eV}, 
    and in the (f) $\Gamma$-M high symmetry plane, obtained with a photon energy of $h\nu = $ \qty{475}{\eV}.
    Measured band dispersions along the R-X and $\Gamma$-M directions are shown in (g) and (h), respectively.
    The features identified in the DFT+$U$calculations are indicated by the corresponding labels in (e)-(h). In (f), a trace of feature 4 from the DFT+$U$ calculation is overlaid as a guide to the eye.
    }
    \label{fig:bigfig}
\end{figure*}   

The DFT calculations of the spin-up Fermi surface in the R-X and $\Gamma$-M high symmetry cuts are shown in Figs. \ref{fig:bigfig}(a) and (b), respectively, corresponding to the red and blue planes in Fig. \ref{fig:scfig}(b).
For the R-X cut in Fig. \ref{fig:bigfig}(a), we identify two main features: a central pocket, denoted as feature 1, and a point-like feature in sixfold coordination, denoted as feature 2. 
Considering the $\Gamma$-M cut in Fig. \ref{fig:bigfig}(b), we denote feature 3 as the central electron-like pocket, and feature 4 as the surrounding triangular features. 
By comparing the Fermi surface cuts to the calculated band dispersion along the R-X and $\Gamma$-M direction in Figs. \ref{fig:bigfig}(c) and (d), features 1 and 3 are identified with the hole-like and electron-like states surrounding the R and $\Gamma$ points, respectively.
We furthermore observe that features 1 and 2 are replicated in the $\Gamma$-M cut, as well as features 3 and 4 in the R-X cut. This is attributed to the unit cell doubling along [111] mapping the $\Gamma$-M plane to the R-X plane, and vice versa. This mapping may also be seen in the fainter band dispersion in Figs. \ref{fig:bigfig}(c) and (d).

The DFT calculated band structure in Figs.~\ref{fig:bigfig}(a)-(d) may be directly compared to the band structure obtained by ARPES, shown in Figs. \ref{fig:bigfig}(e)-(h).
From the photon energy scan in Fig. \ref{fig:kz} (a), $h\nu = $ \qty{369}{\ev} and $h\nu = $ \qty{475}{\ev} are identified as suitable photon energies to approximately probe the R-X and $\Gamma$-M planes, respectively.
In Fig. \ref{fig:bigfig}(e) a Fermi surface cut through the R-X plane is shown.
We identify the hole-like R-pocket (feature 1), the six-fold coordination of the R-points in the second Brillouin zone, and the point-like features (feature 2) as predicted by DFT calculations in Fig. \ref{fig:bigfig}(a).
For the $\Gamma$-M plane, depicted in Fig. \ref{fig:bigfig}(f), we similarly observe the central electron-like pocket (feature 3) and the edges of the surrounding feature 4. 
Beyond this, however, the Fermi surface cut as measured by ARPES shows a trifold modulation that is not present in the calculated Fermi surface. 
This is attributed to a modulation of the photoemission intensity due to the trifold symmetry along the [111] axis.

Figures \ref{fig:bigfig}(g) and (h) show the valence band dispersions along the R-X and $\Gamma$-M directions, respectively. 
As with the Fermi surface cuts in Figs. \ref{fig:bigfig}(e), (f), we observe features 1 and 3 in good agreement with the theoretical band dispersions. However, the dome-like band along the $\Gamma$-M direction predicted by DFT in Fig. \ref{fig:bigfig}(d) is not discernible in the ARPES measurement of Fig. \ref{fig:bigfig}(h).
Furthermore, the backfolded bands due to unit cell doubling are not visible in neither the measured Fermi surface cuts nor the band dispersions. 
In sum, all features present in the experimental band structure measurements are accurately reproduced by the DFT+$U$ calculations, while some features predicted in the DFT+$U$ calculations, such as the backfolded bands, are not observed in the ARPES measurements.

\subsection{Magnetic circular dichroism in resonant ARPES}

\begin{figure}[h]
    \centering
    \includegraphics[width = \linewidth]{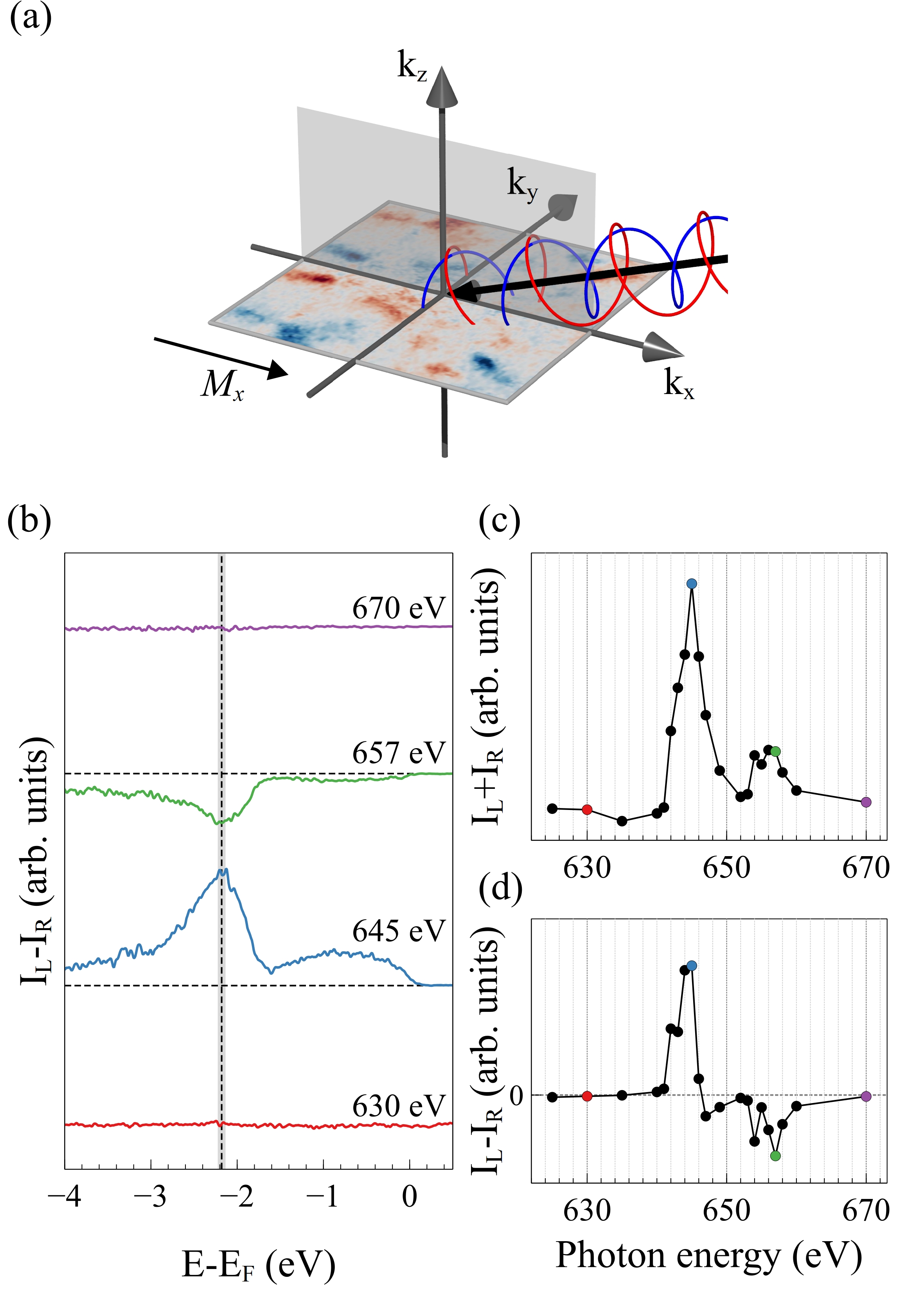}
    \caption{
    (a) Experimental geometry of the measurements probing the circular dichroism in ARPES. The photons, either left or right circularly polarized, impinge on the sample along the black arrow at an incidence angle of 20$^{\circ}$. The plane of light incidence is in the \textit{xz} plane, shown in gray, which coincides with the mirror plane along the $[11\bar{2}]$ crystallographic axis.
    In this geometry, a magnetization $M_x$ is expected to induce magnetic circular dichroism. 
    (b) Angle-integrated CD ($I_{CD}=I_L-I_R$) in photoemission measured at photon energies before, on, and after the Mn $L_3$ and $L_2$ resonances.
    (c) Angle-integrated photoemission intensity ($I=I_L + I_R$) as a function of photon energy at $E-E_F = $ \qty{-2.18}{\ev} integrated over a range of \qty{0.1}{\ev}, as indicated by the vertical dashed line and shaded area in (b).
    (d) Angle-integrated CD as a function of photon energy at $E-E_F = $ \qty{-2.18}{\ev} integrated over a range of \qty{0.1}{\ev}.
    The photon energies depicted in (b) are highlighted by their respective colors in (c) and (d). 
    }
    \label{fig:xmcd_valence}
\end{figure}

We explored circular dichroism (CD) in resonant ARPES at the Mn $L$ X-ray absorption edge as a momentum-resolved probe of spin polarization in the electronic band structure \cite{Kruger2025CircularAltermagnets,Sagehashi2023TheorySurfaces}. Conceptually, the approach can be viewed from two sides: (i) as a regular CD-ARPES experiment but performed in resonance or (ii) as an X-ray magnetic circular dichroism (XMCD) experiment but with additional momentum resolution. 

We start the discussion with a few general considerations. The CD-ARPES measurements were performed with the plane of light incidence (\textit{xz}) aligned with a crystallographic mirror plane $[11\bar{2}]$ along $k_y = 0$, as indicated in Fig.~\ref{fig:xmcd_valence}(a). 
The CD signal is obtained as the difference between photoemission intensities measured with left- and right circularly polarized photons, $I_{CD} = I_L - I_R$ \cite{Oh2025InterplayChirality}. In the absence of magnetic ordering, the crystallographic mirror plane imposes an antisymmetric CD angular distribution $I_{CD}(k_y) = -I_{CD}(-k_y)$ \cite{Figgemeier2025ImagingSpace}, resulting in a net zero angle-integrated CD signal. Note that the latter is also enforced by time-reversal symmetry even for incidence off the mirror plane.
The symmetry-breaking by ferromagnetic order lifts these requirements, allowing for non-zero net CD in photoemission, known as magnetic circular dichroism (MCD) in photoemission \cite{Kuch2001MagneticPhotoemission,Rosenberger2025SpinInterface}. The appearance of MCD, however, relies on spin-orbit interaction and usually is a small effect. Spin-orbit interaction in the photoexcitation process can be enhanced by tuning the photon energy to a resonant absorption edge, leading to pronounced MCD in angle-integrated resonant photoemission \cite{Tjeng1993MagneticNickel,Goedkoop1999MagneticPhotoemission}. This effect is closely related to the XMCD technique, the main difference being that in XMCD, typically the full photoexcitation cross section is monitored (total electron yield). MCD and resonant photoemission may then be combined with angle-resolved photoelectron detection, yielding MCD in resonant ARPES, which naturally combines local magnetic and itinerant band-structure information \cite{DaPieve2016FingerprintsPhotoemission} but to our knowledge has rarely been explored experimentally.

We first consider angle-integrated resonant photoemission experiments for LSMO(111). Figures~\ref{fig:xmcd_valence}(b)-(d) show the photoemission intensity within the plane of light incidence, integrated over an angular range of $\pm15^{\circ}$ around normal emission. As the photon energy is tuned across the Mn $L_{3,2}$ absorption edge, the photoemission intensity is amplified by resonant enhancement \cite{Weinelt1997ResonantEffects}, as seen in Fig.~4(c). On resonance, we observe a pronounced MCD that reverses sign between the $L_{3}$ and $L_{2}$ edges, while only a negligible MCD is found off resonance.
The photon-energy dependence of the MCD signal in Fig.~\ref{fig:xmcd_valence}(d) resembles the Mn $L_{3,2}$ XMCD signal in LSMO(111) measured in total electron yield mode (see supplemental material \cite{supplemental} or Ref. \cite{Zakharova2021InterplayManganites}).
These findings confirm that the observed MCD in resonant photoemission is dominated by the XMCD selection rules \cite{StohrJ1999ExploringSpectroscopy}, in agreement with recent theory \cite{Kruger2025CircularAltermagnets}, and may be regarded as a partial electron yield XMCD signal. 
Therefore, given the grazing-incidence experimental geometry, see Fig.~\ref{fig:xmcd_valence}(a), the observed MCD can be attributed to a magnetization $M_x$ along the $[11\bar{2}]$ axis. 
This remanent magnetization was established without applying an external magnetic field, and appeared to be uniform across the sample. 
The $[11\bar{2}]$ axis is parallel to the step-edges of the thin film. Such a preferential magnetization along the step-edges is consistent with previous reports for LSMO thin films in both the (001) and (111) orientations \cite{Taniuchi2006ObservationMicroscopy, Hallsteinsen2015CrystallineFilms}.
The magnetic origin of the MCD was confirmed by magnetizing the sample along $\pm M_x$, which induced a sign reversal in the MCD signal (see Supplemental Material \cite{supplemental}).
As seen in Fig.~\ref{fig:xmcd_valence}(b), significant MCD is observed across the entire valence band. 
Overall, the data in Fig.~\ref{fig:xmcd_valence} confirm a ferromagnetic state with in-plane anisotropy in the LSMO(111) film.

\begin{figure} 
    \centering
    \includegraphics[width=\linewidth]{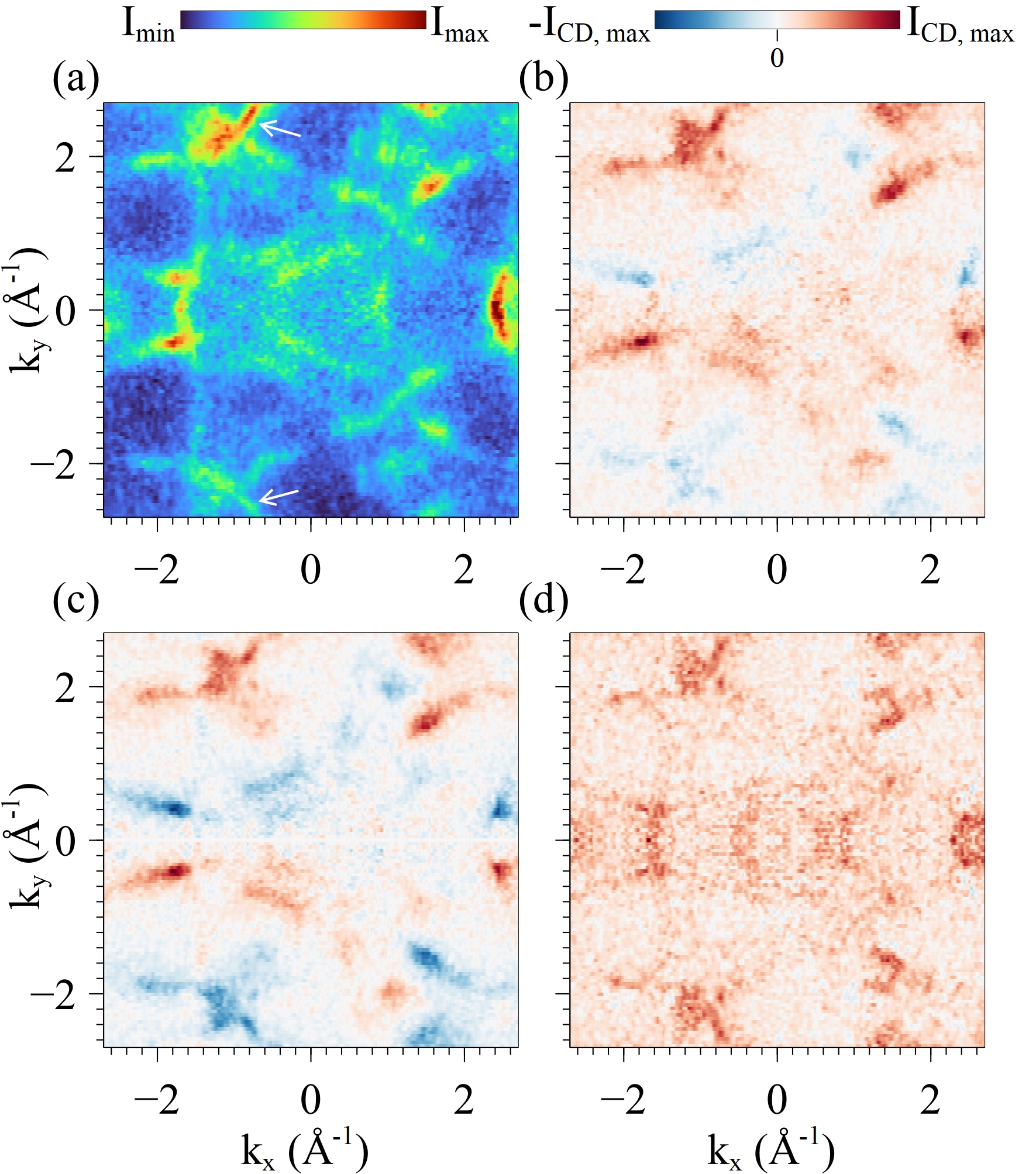}
    \caption{
    ARPES data measured at the Fermi level ($E-E_F = 0$) with a photon energy $h\nu = $ \qty{645}{\eV} tuned to the Mn $L_3$ resonance (cf. Fig. 4). (a) Total intensity obtained by taking the sum of left and right circularly polarized light, $I=I_L+I_R$. (b) Circular dichroism obtained by taking the difference between left and right circularly polarized light, $I_{CD}=I_L-I_R$.  (c) Anti-symmetric contribution to $I_{CD}$ extracted from (b), representing the geometrically induced circular dichroism. (d) Symmetric contribution to $I_{CD}$ extracted from (b), representing the magnetically induced circular dichroism.
    }
    \label{fig:cd}
\end{figure}

Figure 5 presents measurements of the MCD in resonant ARPES. Notably, the momentum distribution of the resonant photoemission intensity in Fig.~5(a) shows well-defined, sharp features, in agreement with previous work \cite{Lev2015FermiMagnetoresistance}. This demonstrates that the core-level absorption and subsequent participator Auger decay, which dominate the resonant photoemission intensity, form a momentum-selective process that probes itinerant Bloch states. The magnetization $M_x$ breaks the mirror symmetry of the experimental geometry, allowing for deviations from the relations $I(k_y) = I(-k_y)$ and $I_{CD}(k_y) = -I_{CD}(-k_y)$ that are enforced for the intensity $I= I_L +I_R $ and the circular dichroism $I_{CD}= I_L -I_R $ in a fully mirror symmetric situation \cite{PhysRevLett.132.196402}.
Indeed, the experimental data for both $I$ and $I_{CD}$ in Figs.~5(a)-(b) show clear signatures of broken mirror symmetry. The intensity distribution in Fig.~\ref{fig:cd}(a) shows higher intensities at positive $k_y$ for most features, especially those at larger $k_y$, with some exceptions, where the intensity is higher at negative $k_y$. See for instance the features indicated by the white arrows.
The CD momentum distribution in Fig.~5(b) shows marked deviations from a perfect anti-symmetry, with positive/red contributions outweighing negative/blue contributions. We note that for excitation energies off resonance, we observed mirror-symmetry preserving momentum distributions for the intensity and the circular dichroism within experimental uncertainty (see Supplemental Material \cite{supplemental}), which indicates that the non-resonant MCD \cite{Kuch2001MagneticPhotoemission} is weak in LSMO, at least in the soft X-ray regime. This finding also helps to rule out potential misalignment in the experimental setup, further supporting the magnetic origin of the observed mirror-symmetry breaking. 

The net MCD in the angle-integrated data in Fig.~\ref{fig:xmcd_valence} manifests in the momentum-resolved resonant CD data in Fig. 5 as a contribution that is symmetric with respect to the crystallographic mirror $xz$ plane. We extract the anti-symmetric and symmetric contributions to the total circular dichroism and plot them in Figs. 5(c) and (d), respectively. The symmetric contribution constitutes the momentum distribution of the resonant MCD. 
While there is some variation in the size of the MCD for different features in the Fermi surface cut, the MCD shows a momentum structure that is largely inherited from the intensity distribution in Fig. \ref{fig:cd}(a).

Our experiments for LSMO(111) establish a sizable momentum-dependent MCD in resonant ARPES at the Mn $L_{3,2}$ Mn absorption edge. 
The approach combines the spin sensitivity of XMCD with the momentum sensitivity of ARPES. Such spectroscopic information is otherwise accessible only through spin-resolved ARPES, a methodology that remains experimentally challenging, particularly for complex momentum distributions \cite{Tan2025ExchangeHalf-Metal,Chen2026MomentumApplications}. Our results support recent theory \cite{Kruger2025CircularAltermagnets}, which proposes MCD in resonant ARPES as a probe of unconventional magnetic states with even- or odd-parity spin momentum textures, such as altermagnetism \cite{Smejkal2022EmergingAltermagnetism} or p-wave magnetism \cite{Yamada2025AHelix}. 

\section{Conclusions}

In conclusion, we have explored the electronic structure of a (111)-oriented LSMO thin film through soft X-ray angle-resolved photoemission spectroscopy. By exploiting the intrinsically sharp definition of the out-of-plane momentum in the soft X-ray photon-energy regime, we probe the three-dimensional bulk electronic structure of LSMO. The measured electronic structure is compared to the results of first-principles calculations performed on an LSMO supercell generated through the SQS method, largely in agreement with the measured electronic structure. 

Moreover, we have investigated the circular dichroism in on- and off-resonant ARPES, and find a significant, momentum-dependent MCD at the Mn $L_{3,2}$ resonance that is negligible at off-resonant energies. 
Combining spin- and momentum selectivity may enable new types of experiments probing unconventional magnetic materials \cite{Kruger2025CircularAltermagnets}.

\begin{acknowledgements}
\section{Acknowledgments}
The authors acknowledge funding through the Research Council of Norway on the HONEYCOMB project (325063) and through the projects 323766, 302506, 354614, and 301954, and its Centres of Excellence funding scheme Grant No. 262633 “QuSpin.” F. G. was supported by the German Research Foundation (Deutsche Forschungsgemeinschaft, DFG) through Grant No. 556350547. Computational resources were provided by Sigma2 under the project NN9264K. We acknowledge DESY (Hamburg, Germany), a member of the Helmholtz Association HGF, for the provision of experimental facilities. Parts of this research were carried out at PETRA III. 
Data were collected using the ASPHERE III instrument at beamline P04. Beamtime was allocated for proposal II-20230019. The ASPHERE III instrument is funded by the ErUM-Pro program (grant number 05K25FK4) of the German Federal Ministry of Research, Technology and Space (BMFTR). 
We would like to thank Frank Scholz and Jörn Seltmann for assistance in using beamline P04. 

\end{acknowledgements}

\bibliography{ref}
\end{document}


\maketitle

\section{Sample growth and characterization}

\subsection{Sample growth}
The La$_{0.7}$Sr$_{0.3}$MnO$_3$ thin film was grown by pulsed laser deposition (PLD) on (111)-oriented SrTiO$_3$ substrates (Shinkosha). Before growth, the substrate was cleaned in subsequent ultrasonic baths of acetone, ethanol and deionized water for 5 minutes each at room temperature followed by etching in a buffered hydrofluoric acid solution with a 1:9 ratio of HF to NH$_4$F for 1 minute. The substrate was annealed at 1050 $^\circ$C under oxygen flow for 1 hour. Finally, once inserted into the PLD chamber, the substrate was annealed at 280 $^\circ$C for 1 hour in a oxygen partial pressure of \qty{0.3}{\mbar} to remove organic contaminants from the surface.
A KrF excimer laser with wavelength $\lambda$ = 248 nm was used to ablate from a La$_{0.7}$Sr$_{0.3}$MnO$_3$ target (Kurt J. Lesker) at a target-substrate distance of 40 mm with a laser fluence of \qty{1.2}{\fluence}. The laser frequency was set to \qty{2}{\hertz}.
During growth, the sample was kept at 700 $^\circ$C with an oxygen background pressure of \qty{0.12}{\mbar}. After growth, the sample was annealed for 1 hour at 600 $^\circ$C in \qty{300}{\mbar} O$_2$ partial pressure.

\begin{figure}
    \centering
    \includegraphics[width=\linewidth]{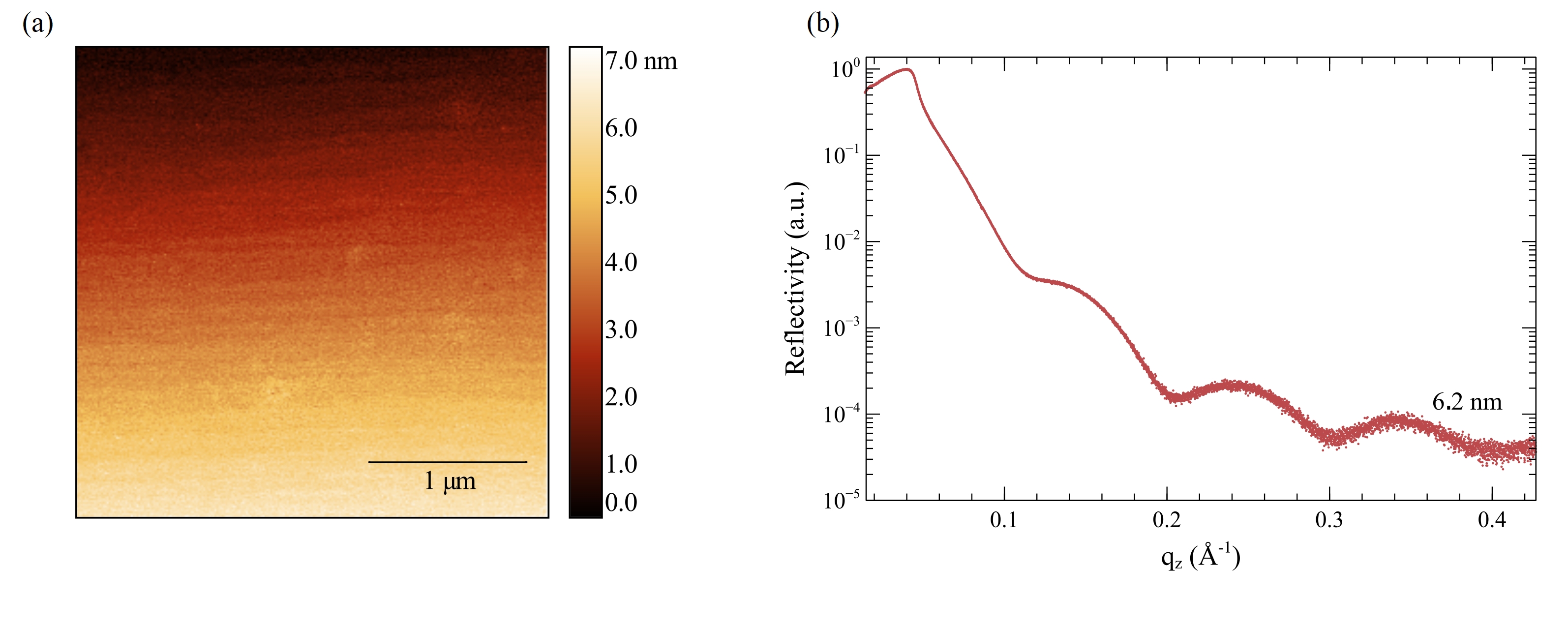}
    \caption{
    (a) AFM image of the LSMO thin film, showing a step-and-terrace structure. (b) X-ray-reflectivity profile of the LSMO thin film, with thickness fringes corresponding to a thickness of \qty{6.2}{\nm}. 
    }
    \label{fig:xrr}
\end{figure}

After growth, atomic force microscopy (AFM) images, shown in Fig. \ref{fig:xrr} (a), shows a smooth surface with step-terraces inherited from the SrTiO$_3$ substrate. A X-ray reflectivity profile was taken to determine the film thickness. This is shown in Fig. \ref{fig:xrr} (b), where the thickness fringes correspond to a film thickness of \qty{6.2}{\nm}.

\subsection{Surface preparation}
\begin{figure}
    \centering
    \includegraphics[width=\linewidth]{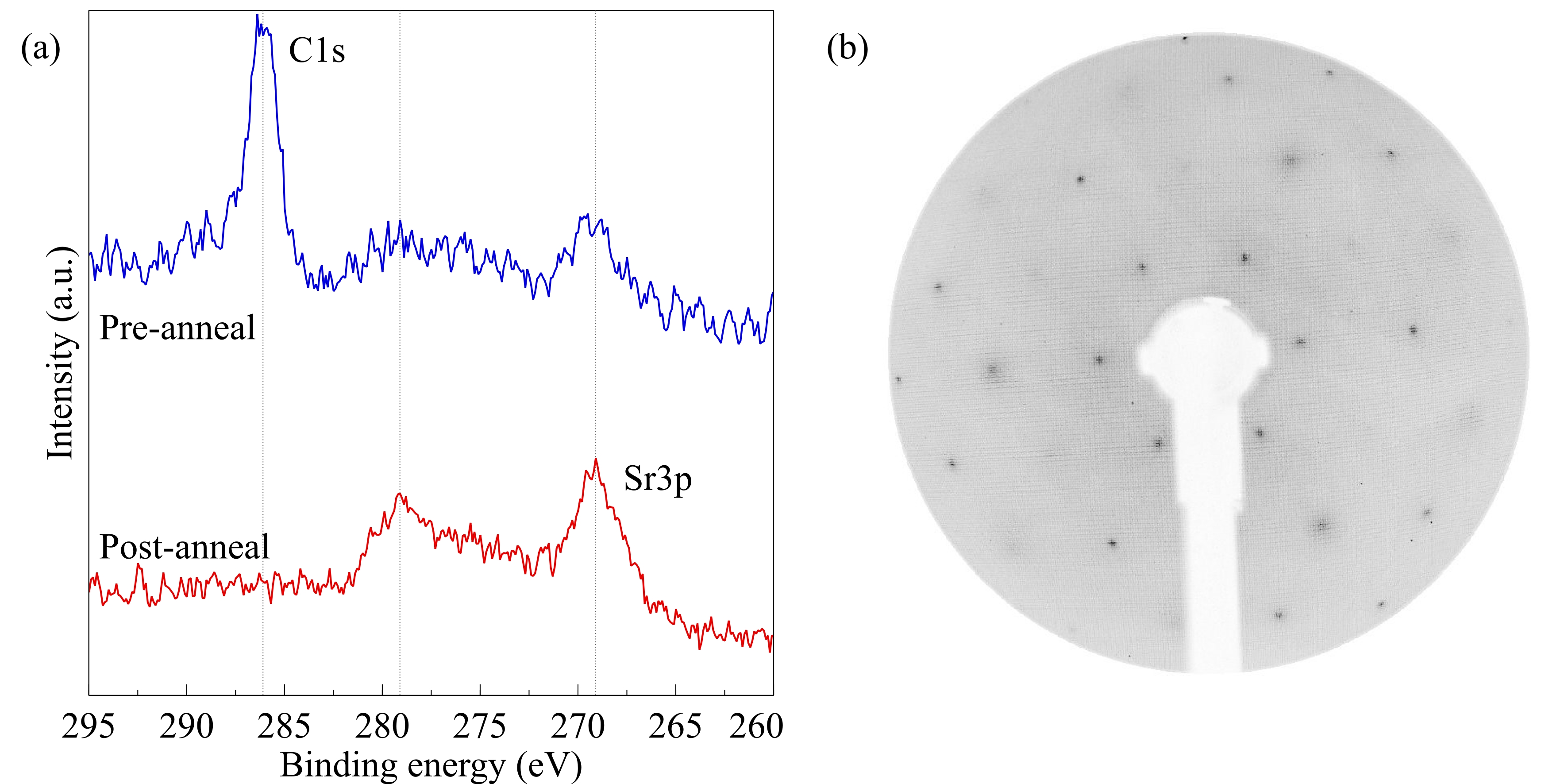}
    \caption{
    (a) XPS spectra of C 1s and Sr  3p peaks before (red) and after (blue) the surface preparation procedure, indicating the removal of C contamination and the emergence of Sr peaks. 
    (b) LEED diffractogram taken after the preparation procedure with incident electrons with an energy of \qty{156}{\ev}, demonstrating that a well-ordered sample surface has been obtained.
    }
    \label{fig:xps_leed}
\end{figure}

The sample was transported through air between the growth and the conducted ARPES measurements. As such, a surface preparation procedure was necessary upon reintroduction to vacuum. The procedure involved annealing the sample at temperature of $\sim$ \qty{640}{\kelvin} in O$_2$ gas with a partial pressure of \qty{1e-5}{\mbar}. 
The efficacy of the preparation procedure was evaluated by X-ray photoelectron spectroscopy (XPS) and low energy electron diffraction (LEED). 
Fig. \ref{fig:xps_leed} (a) includes two XPS spectra from before and after the annealing procedure, respectively, and demonstrates the successful removal of C contamination and the emergence of the Sr 3p peaks. The XPS spectra were obtained using a Al K$\alpha$ source with photon energy $h\nu = $\qty{1486.6}{\ev}.
In Fig. \ref{fig:xps_leed} (b), a LEED diffractogram taken after the preparation procedure indicates that a highly crystalline sample surface was obtained.

\section{Effect of octahedral tilt pattern on calculated band structures}

\begin{figure}
    \centering
    \includegraphics[width=\linewidth]{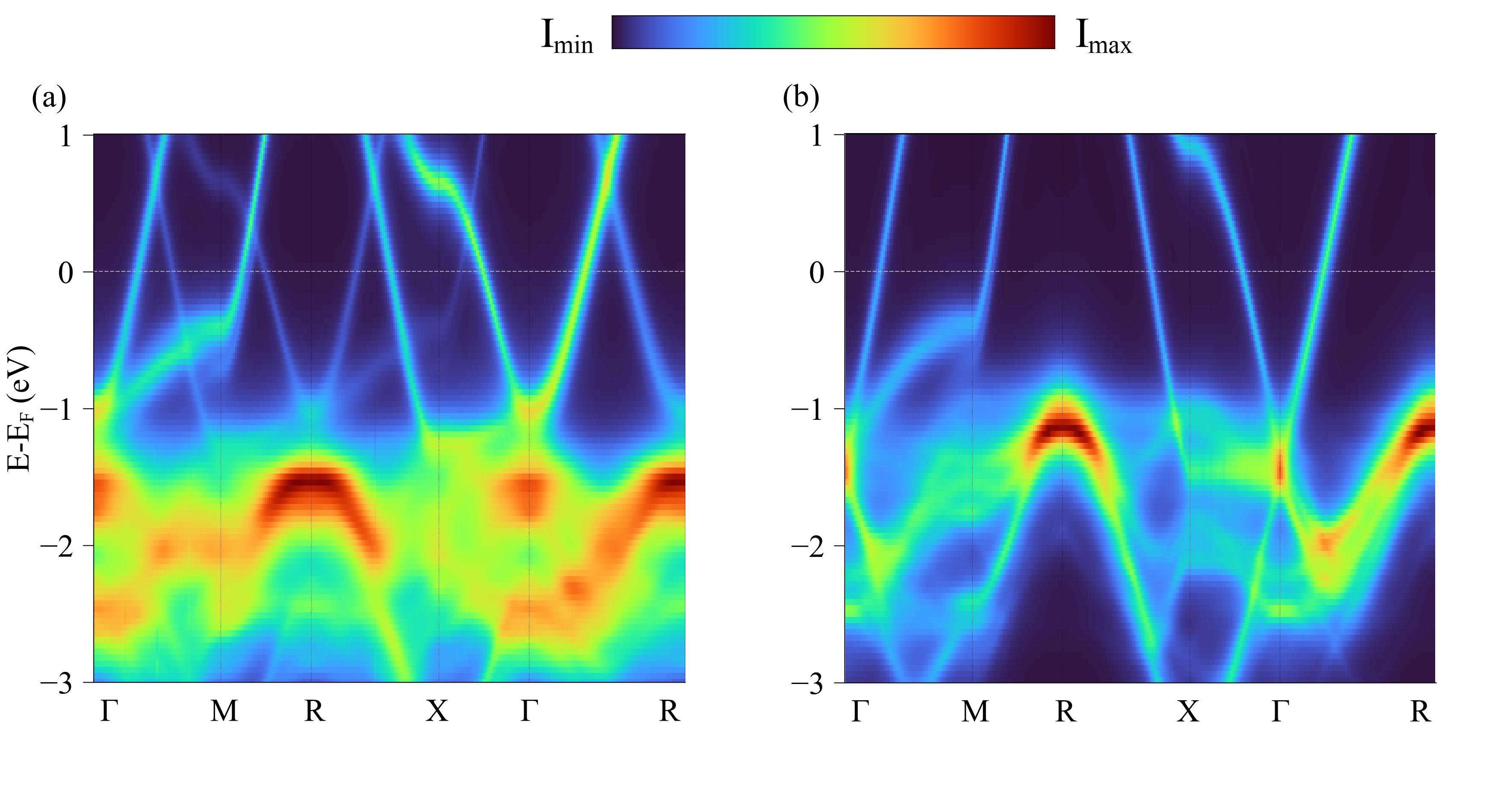}
    \caption{
    Spin up electronic band structures calculated by DFT+$U$ for a supercell with (a) $a^-a^-a^-$ octahedral tilt pattern as expected for the $R\bar{3}c$ space group and (b) for a supercell with no octahedral tilting.
    }
    \label{fig:tiltcomp}
\end{figure}

After optimizing the supercell geometry, the tilt of the O octahedra follows a $a^+b^-b^-$ pattern 
, such that the tilt along one axis is in-phase with the remaining two axes anti-phase. 
This deviates from the $a^-a^-a^-$ pattern in bulk LSMO, in which all axes exhibit anti-phase tilt patterns.
The deviation is attributed to a small energy difference between the $a^+b^-b^-$ and $a^-a^-a^-$ geometries.
%
In order to understand the impact of the deviating octahedral tilt, we here include calculated band structures for supercells in non-optimized geometries corresponding to the $R\bar{3}c$ and $Pm\bar{3}m$ space groups, respectively.
In the case of the $R\bar{3}c$ space group, the octahedral tilt pattern is $a^-a^-a^-$, meaning anti-phase tilting of equal magnitude along all cubic axes. The resulting band structure is shown in Fig. \ref{fig:tiltcomp} (a). we observe no significant differences from the band structure obtained for the supercell in an optimized geometry as shown in the main text.
For the $Pm\bar{3}m$ supercell, there is no octahedral tilt. We observe that, as opposed to the supercells with octahedral tilt, the calculated band structure exhibits no mapping of the R-X plane to the $\Gamma$-M plane, indicating that this effect is indeed a result of the unit cell doubling along the [111] axis in the presence of octahedral tilt.


\section{Circular dichroism}
\begin{figure}
    \centering
    \includegraphics[width=\linewidth]{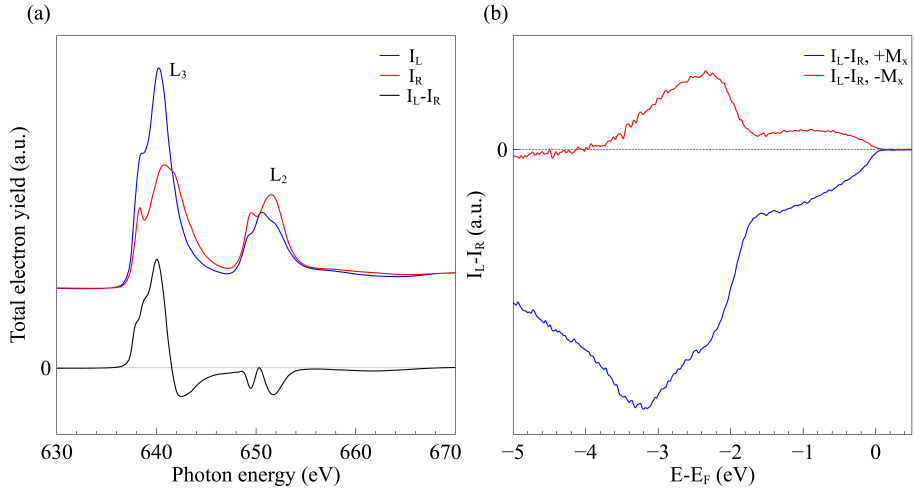}
    \caption{
    (a) XAS and XMCD spectra across the Mn L$_3$ and L$_2$ edges, indicating an in-plane ferromagnetic ordering in the sample.
    (b) Angle-integrated photoemission CD measured at the L$_3$ resonance for opposite magnetization directions $\pm M_x$ along the plane of light incidence. Switching the direction of the magnetization clearly switches the sign of the photoemission CD.
    }
    \label{fig:xmcd}
\end{figure}

In Fig. 4 (d) in the main text, we compare the difference between the angle-integrated photoemission intensity for left and right circularly polarized light to the XMCD spectrum of LSMO. To see this clearly, consider the XMCD spectrum of the sample as shown in Fig. \ref{fig:xmcd} (a)
The XMCD spectrum exhibits the same step-wise peak at the L$_3$ resonance and double peak at the L$_2$ resonance, along with the sign change between the two peaks. This indicates an in-plane ferromagnetic ordering in the sample.

To confirm the magnetic origin of the MCD discussed in the main text, we measured the photoemission CD for magnetization along $\pm M_x$, depicted in Fig. \ref{fig:xmcd} (b). 
To switch the magnetization, the sample was magnetized \textit{in-situ} at approximately \qty{30}{\kelvin} using two oppositely oriented permanent magnets along \textit{x}.
When reversing the direction of the magnetization $M_x$, the sign of the angle-integrated photoemission CD likewise switches. 
The difference in the CD intensity between $\pm M_x$ is attributed to inconsistencies in the magnetization procedure.

\begin{figure}
    \centering
    \includegraphics[width=\linewidth]{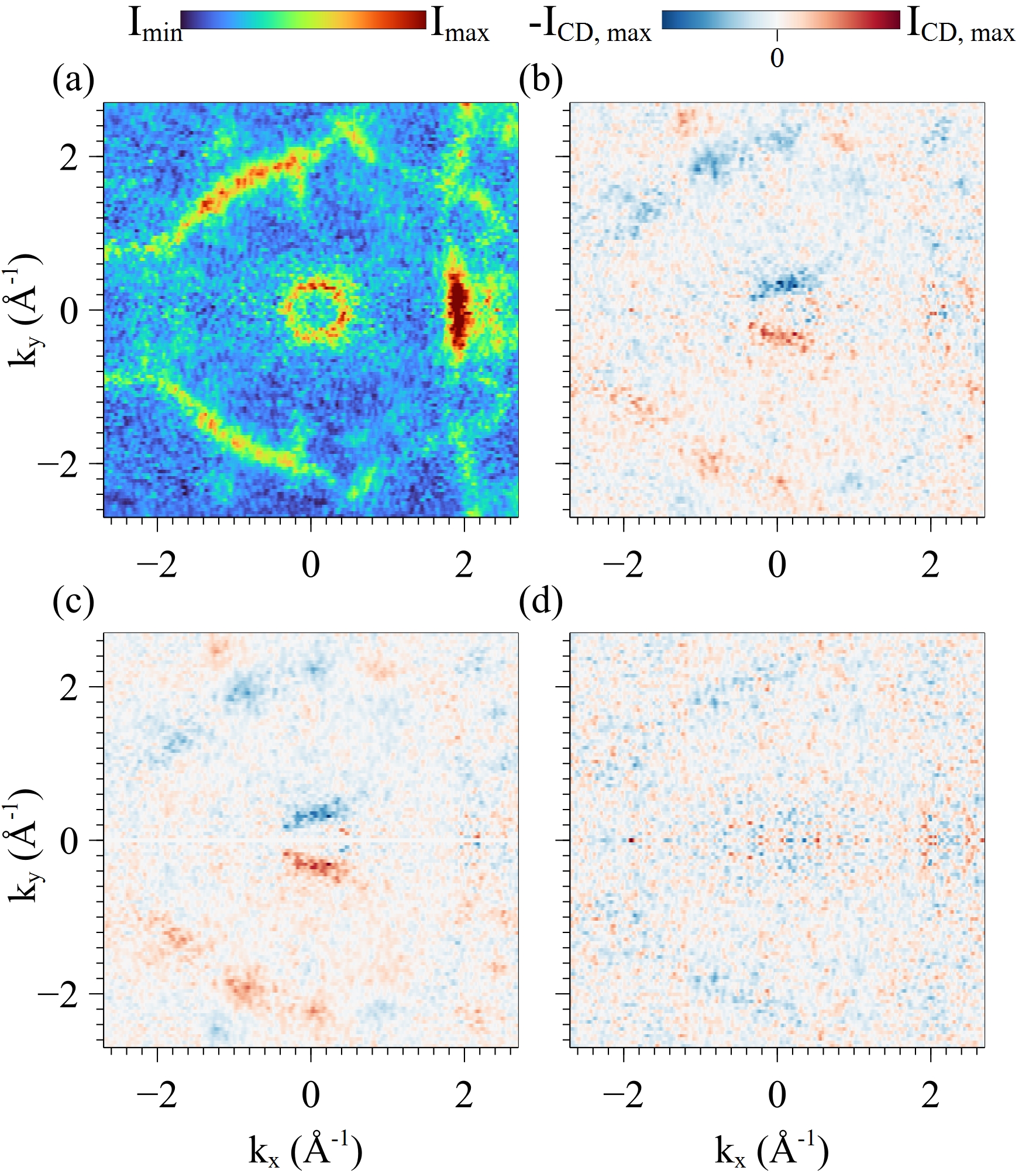}
    \caption{
    ARPES data measured at the Fermi level with $h\nu = $ \qty{475}{\ev}.
    (a) Total intensity obtained as the sum of left and right circularly polarized light, $I = I_L + I_R$.
    (b) Circular dichroism obtained as the difference between left and right circularly polarized light $I_{CD} = I_L - I_R$.
    (c) Anti-symmetric contribution extracted from (b).
    (d) Symmetric contribution extracted from (b).
    }
    \label{fig:cd475}
\end{figure}

The main text discusses the momentum-resolved magnetic circular dichroism (MCD) in resonant ARPES, obtained using a photon energy of $h\nu = $ \qty{645}{\ev}. 
To explore the momentum distribution of the circular dichroism off resonance, we here include ARPES data measured with a photon energy of $h\nu = $ \qty{475}{\ev}. 
In the integrated intensity $I = I_L + I_R$, shown in Fig. \ref{fig:cd475} (a), we observe a momentum distribution that appears mirror symmetric with respect to $k_y = 0$. Considering the circular dichroism $I_{CD} = I_L- I_R$, show in Fig. \ref{fig:cd475} (b), we observe an anti-symmetric distribution. As discussed in the main text, this is in agreement with the distribution required by the crystallographic mirror plane along $k_y = 0$ in the absence of magnetic ordering. To confirm this, we extract the anti-symmetric and symmetric contributions to the total circular dichroism, shown in Figs. \ref{fig:cd475} (c) and (d), respectively. In contrast to the resonant MCD shown in Fig.~5 of the main text, we here observe that the total circular dichroism is dominated by the anti-symmetric contribution, with the symmetric contribution being negligible.

